\documentclass[conference]{IEEEtran}
\usepackage{amssymb}
\usepackage{cite}

\usepackage{multirow}
\usepackage{subfig}
\usepackage{graphicx}
\usepackage[export]{adjustbox}
\usepackage{algpseudocode,algorithmicx}
\usepackage[linesnumbered,ruled]{algorithm2e}
\usepackage{amsmath}
\usepackage{listings}
\usepackage{color}
\usepackage{float}
\usepackage{tikz}
\usepackage{url}
\usepackage{microtype}
\definecolor{dkgreen}{rgb}{0,0.6,0}
\definecolor{gray}{rgb}{0.5,0.5,0.5}
\definecolor{mauve}{rgb}{0.58,0,0.82}

\lstdefinestyle{customasm}{
  belowcaptionskip=1\baselineskip,
  xleftmargin=\parindent,
  language=[x86masm]Assembler,
  basicstyle=\footnotesize\ttfamily,
  commentstyle=\itshape\color{purple!40!black},
}
\begin{document}
%
\title{Convolutional Neural Networks over Control Flow Graphs for Software Defect Prediction}

\author{\IEEEauthorblockN{Anh Viet Phan, Minh Le Nguyen}
\IEEEauthorblockA{Japan Advanced Institute of Information Technology\\Nomi, Japan 923-1211\\
Email: \{anhphanviet,nguyenml\}@jaist.ac.jp}
\and
\IEEEauthorblockN{Lam Thu Bui}
\IEEEauthorblockA{Le Quy Don Technical University\\Ha Noi, Vietnam\\
Email: lambt@lqdtu.edu.vn}
}
\maketitle

\begin{abstract}
Existing defects in software components is unavoidable and leads to not only a waste of time and money but also many serious consequences. To build predictive models, previous studies focus on manually extracting features or using tree representations of programs, and exploiting different machine learning algorithms. However, the performance of the models is not high since the existing features and tree structures often fail to capture the semantics of programs.       
To explore deeply programs' semantics, this paper proposes to leverage precise graphs representing program execution flows, and deep neural networks for automatically learning defect features. Firstly, control flow graphs are constructed from the assembly instructions obtained by compiling source code; we thereafter apply multi-view multi-layer directed graph-based convolutional neural networks (DGCNNs) to learn semantic features. The experiments on four real-world datasets show that our method significantly outperforms the baselines including several other deep learning approaches.
\end{abstract}

\begin{IEEEkeywords}
Software Defect Prediction, Control Flow Graphs, Convolutional Neural Networks
\end{IEEEkeywords}


%
\IEEEpeerreviewmaketitle
\section{Introduction}
\label{sec:introduction}
Software defect prediction has been one of the most attractive research topics in the field of software engineering. According to regular reports, semantic bugs result in an increase of application costs, trouble to users, and even serious consequences during deployment. Thus, localizing and fixing defects in early stages of software development are urgent requirements. Various approaches have been proposed to construct models which are able to predict whether a source code contains defects. These studies can be divided into two directions: one is applying machine learning techniques on data of software metrics that are manually designed to extract features from source code; the other is using programs' tree representations and deep learning to automatically learn defect features.   

The traditional methods focus on designing and combining features of programs. For product metrics, most of them are based on statistics on source code. For example, Halstead metrics are computed from numbers of operators and operands~\cite{halstead1977elements}; CK metrics are measured by function and inheritance counts ~\cite{chidamber1994metrics}; McCabe's metric estimates the complexity of a program by analyzing its control flow graph~\cite{mccabe1976complexity}. However, according to many studies and surveys, the existing metrics often fail in capturing the semantics of programs~\cite{jones1997strengths,menzies2007data}. As a result, although many efforts have been made such as adopting robust learning algorithms and refining the data, the classifier performance is not so high~\cite{catal2011software}.     

Recently, several software engineering problems have been successfully solved by exploiting tree representations of programs - the Abstract Syntax Trees (ASTs)~\cite{louden2011programming}. In the field of machine learning, the quality of input data directly affects the performance of learners. Regarding this, due to containing rich information of programs, tree-based approaches have shown significant improvements in comparison with previous research, especially software metrics-based. Mou et al. proposed a tree-based convolutional neural network to extract structural information of ASTs for classifying programs by functionalities~\cite{mou2016convolutional}. Wang et al. employed a deep belief network to automatically learn semantic features from AST tokens for defect prediction~\cite{wang2016automatically}. Kikuchi et. al measured the similarities between tree structures for source code plagiarism detection~\cite{kikuchi2014source}.
\begin{figure}[t]
  \centering
  \subfloat[File 1.c]{\includegraphics[width=4.2cm,valign=t]{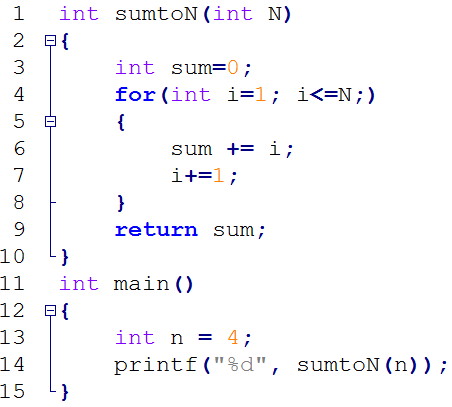}\label{fig:c_code1}}
  \hfill
  \subfloat[File 2.c]{\includegraphics[width=4.2cm,valign=t]{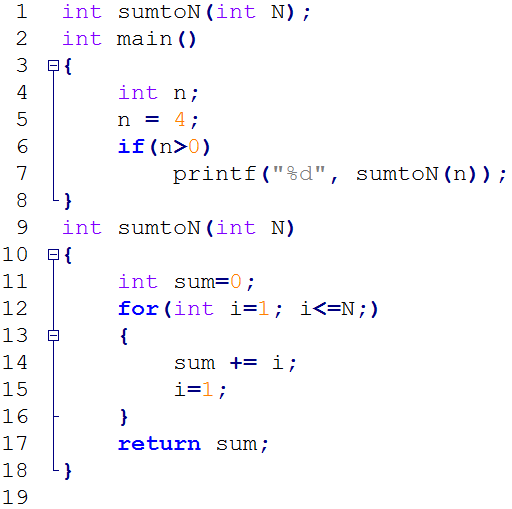}\label{fig:c_code2}}
  \caption{A motivating example}
  \label{fig:two_ccode_examples}
\end{figure}

However, defect characteristics are deeply hidden in programs' semantics and they only cause unexpected output in specific conditions~\cite{white2015toward}. Meanwhile, ASTs do not show the execution process of programs; instead, they simply represent the abstract syntactic structure of source code. Therefore, both software metrics and AST features may not reveal many types of defects in programs. For example, we consider the procedures with the same name \texttt{sumtoN} in two \texttt{C} files \texttt{File 1.c} and \texttt{File 2.c} (Fig.~\ref{fig:two_ccode_examples}). Two procedures have a tiny difference at line 7 \texttt{File 1.c} and line 15 in \texttt{File 2.c}. As can be seen a bug from \texttt{File 2.c}, the statement \texttt{i=1} causes an infinite loop of the \texttt{for} statement in case \texttt{N>=1}. Whereas using RSM tool\footnote{\url{http://msquaredtechnologies.com/m2rsm/}} to extract the traditional metrics, their feature vectors are exactly matching, since two procedures have the same lines of code, programming tokens, etc. Similarly, parsing the procedures into ASTs using Pycparser\footnote{\url{https://pypi.python.org/pypi/pycparser}}, their ASTs are identical. In other words, both metrics-based and tree-based approaches are not able to distinguish these programs.

To explore more deeply programs' semantics, this paper proposes combining precise graphs which represent execution flows of programs called Control Flow Graphs (CFGs), and a powerful graphical neural work. Regarding this, each source code is converted into an execution flow graph by two stages: 1) compiling the source file to the assembly code, 2) generating the CFG from the compiled code. Applying CFGs of assembly code is more beneficial than those of ASTs. Firstly, assembly code contains atomic instructions and their CFGs indicate step-by-step execution process of programs. After compiling two source files (Fig.~\ref{fig:two_ccode_examples}) we observed that \texttt{i+=1} and \texttt{i=1} are translated into different instructions. Secondly, assembly code is refined because they are the products after AST processing of the compiler. The compiler applies many techniques to analyze and optimize the ASTs. For example, two statements \texttt{int n; n=4;} in the main procedure of \texttt{File 2.c} are treated in a similar way with the statement \texttt{int n = 4}; and, the \texttt{if} statement can be removed since the value of \texttt{n} is identified. Meanwhile, ASTs just describe the syntactic structures of programs, and they may contain many redundant branches. Assume that if the statement \texttt{i=1} in procedure \texttt{sumtoN} of \texttt{File 2.c} is changed to \texttt{i+=1} then both programs are the same. However, their AST structures have many differences including the positions of subtrees of two functions \texttt{main} and \texttt{sumtoN}, the separation of declaration and assignment of the variable \texttt{n}, and the function prototype of \texttt{sumtoN} in \texttt{File 2.c}.   

We thereafter leverage a directed graph-based convolutional neural network (DGCNN) on CFGs to automatically learn defect features. The advantage of the DGCNN is that it can treat large-scale graphs and process the complex information of vertices like CFGs. The experimental results on four real-world datasets show that applying DGCNN on CFGs significantly outperforms baselines in terms of the different measures\footnote{The source code and collected datasets are publicly available at \url{https://github.com/nguyenlab/DGCNN}}.              

The main contributions of the paper can be summarized as follows:
\begin{itemize}
\item Proposing an application of a graphical data structure namely Control Flow Graph (CFG) to software defect prediction and experimentally proving that leveraging CFGs is successful in building high-performance classifiers.
\item Presenting an algorithm for constructing Control Flow Graphs of assembly code.
\item Formulating an end-to-end model for software defect prediction, in which a multi-view multi-layer convolutional neural networks is adopted to automatically learn the defect features from CFGs.
\item The model implementation is released to motivate related studies.  
\end{itemize}
The remainder of the paper is organized as follows: Section~\ref{sec:approach} explains step-by-step our approach for solving software defect prediction problem from processing data to adapting the learning algorithm. The settings for conducting the experiments such as the datasets, the algorithm hyper-parameters, and evaluation measures are indicated in Section~\ref{sec:experiments}. We analyze experimental results in Section~\ref{sec:results}, and conclude in Section~\ref{sec:conclusions}.   
\section{The Proposed Approach}
\label{sec:approach}
This section formulates a new approach to software defect prediction by applying graph-based convolutional neural networks over control flow graphs of binary codes. Our proposed method is illustrated in Fig.~\ref{fig_gcnn_architecture} includes two steps: 1) generating CFGs which reveal the behavior of programs, and 2) applying a graphical model on CFG datasets. In the first step, to obtain the graph representation of a program, the source code is compiled into an assembly code using g++ on Linux. The CFG thereafter is constructed to describe the execution flows of the assembly instructions. In the second step, we leverage a powerful deep neural network for directed labeled graphs, called the multi-view multi-layer convolutional neural network, to automatically build predictive models based on CFG data.   
\subsection{Control Flow Graphs}
\label{subsec:CFG}
A control flow graph (CFG) is a directed graph, G = (V,E) where V is the set of vertices $\{v_1,v_2,...,v_n\}$ and E is the set of directed edges $\{(v_i,v_j) , (v_k,v_l),...\}$~\cite{allen1970control}. In CFGs, each vertex represents a basic block that is a linear sequence of program instructions having one entry point (the first instruction executed) and one exit point (the last instruction executed); and the directed edges show control flow paths. 

This paper aims to formulate a method for detecting faulty source code written in \texttt{C} language. Regarding this, CFGs of assembly code, the final products after compiling the source code, server as the input to learn faulty features using machine algorithms. Based on recent research, it has been proved that CFGs are successfully applied to various problems including malware analysis~\cite{bruschi2006detecting,anderson2011graph}, software plagiarism~\cite{chae2013software,sun2014detecting}. Since semantic errors are revealed while programs are running, analyzing the execution flows of the assembly instructions may be helpful for distinguishing faulty patterns from non-faulty ones. 
\begin{figure}[!t]
  \centering
  \subfloat[]{\includegraphics[width=4.2cm]{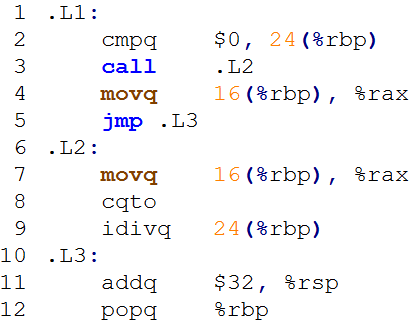}\label{fig:asm_codesample1}}
  \hfill
  \subfloat[]{\includegraphics[width=4cm]{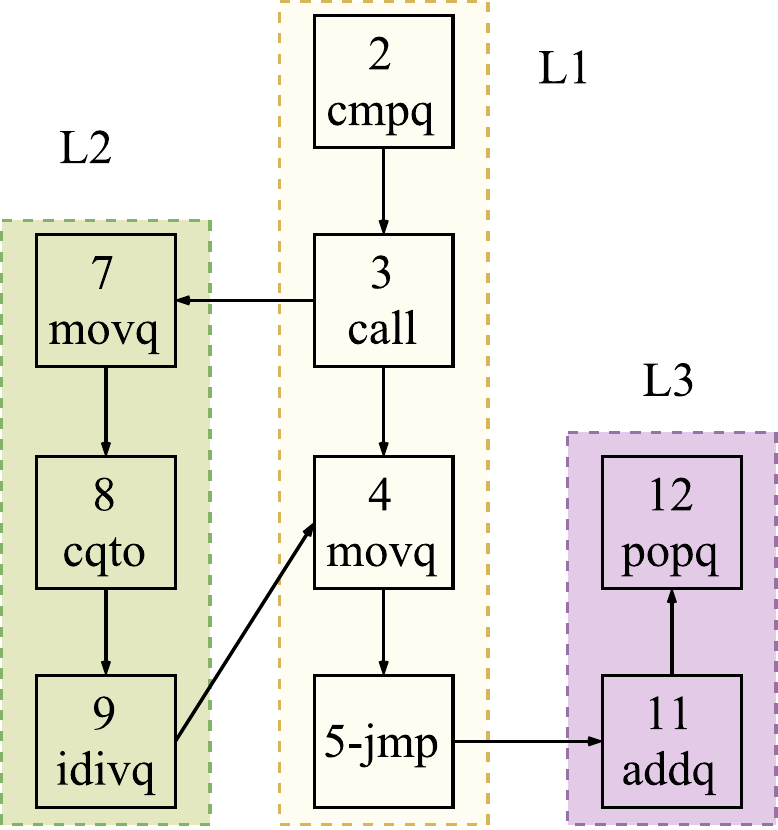}\label{fig:cfg_asm1}}
  \caption{An example of a Control Flow Graph (CFG).(~\ref{fig:asm_codesample1}) a fragment of assembly code;(~\ref{fig:cfg_asm1}) the CFG of the code fragment (each node is viewed by the line number and the name of the instruction).}
  \label{fig:cfg_example}
\end{figure}

Fig.~\ref{fig:cfg_example} illustrates an example of the control flow graph constructed from an assembly code snippet, in which each vertex corresponds to an instruction and a directed edge shows the execution path from an instruction to the other. The pseudo-code to generate CFGs is shown in Algorithm~\ref{algorithm:buildCFG}. The algorithm takes an assembly file as the input, and outputs the CFG. Building the CFG from an assembly code includes two major steps. In the first step, the code is partitioned into blocks of instructions based on the labels (e.g. \texttt{L1, L2, L3} in Fig.~\ref{fig:cfg_example}). The second step is creating the edges to represent the control flow transfers in the program. Specifically, the first line invokes procedure \texttt{initialize\_Blocks} to read the file contents and return all the blocks. In line 2, the set of edges is initially set to empty. From line 3 to 24, the graph edges are created by traversing all instructions of each block and considering possible execution paths from the current instruction to others. For a block, because the instructions are executed in sequence, every node has an outgoing edge to the next one (line 5-9). Additionally, we consider two types of instructions which may have several targets. For \texttt{jump} instructions, an edge is added from the current instruction to the first one of the target block. We use two edges to model function calls, in which one is from the current node to the first instruction of the function and the other is from the final instruction of the function to the next instruction of the current node (line 10-24). Finally, the graphs are formed from the instruction and edge sets (line 25-26).     
\begin{algorithm}
    \SetKwInOut{Input}{Input}
    \SetKwInOut{Output}{Output}
    \Input{$asm\_file$ - A file of assembly code }
    \Output{The graph representation of the code}
      $blocks \gets$ initialize\_Blocks($asm\_file$)\;
      $edges \gets \{\}$\;
        \For{$i\leftarrow0$ \KwTo $\left\vert{blocks}\right\vert$}
        {
        	\For{$j\leftarrow 0$ \KwTo $\left\vert{blocks[i].instructions}\right\vert$}
            {
              \If{$ j > 0$}
              {
					$inst\_1 \gets  blocks[i].instructions[j-1]$\;
                    $inst\_2 \gets  blocks[i].instructions[j]$\;
                    $edges.add$(new\_Edge($inst\_1, inst\_2)$)\;
              }
              \If{inst\_1.type=``jump'' or inst\_1.type=``call''}
              {
              	 $label \gets inst\_1.params[0]$\;
                 $to\_block \gets $find\_Block\_by\_Label($label$)\;
                 \If {to\_block $\neq$ NULL}
                 {
                   $inst\_2 \gets  to\_block.first\_instruction$\;
                   $edges.add$(new\_Edge($inst\_1, inst\_2$))\;
                   \If{inst\_1.type=``call''}
                   {
                      $inst\_2 \gets  to\_block.last\_instruction$\;
                      $inst\_1 \gets  inst\_1.next$\;
                      $edges.add$(new\_Edge($inst\_2, inst\_1$))\;
                   }
                 }
              }
            }
        }
       $instructions \gets$ get\_All\_Instructions($blocks$)\;   
       \Return {construct\_Graph(instructions, edges)}\;
    \caption{The algorithm for constructing Control Flow Graphs from assembly code}
    \label{algorithm:buildCFG}
\end{algorithm}

\subsection{Multi-view Multi-layer Graph-based Convolutional Neural Networks}

\begin{figure*}
\centering
\includegraphics[width=18cm]{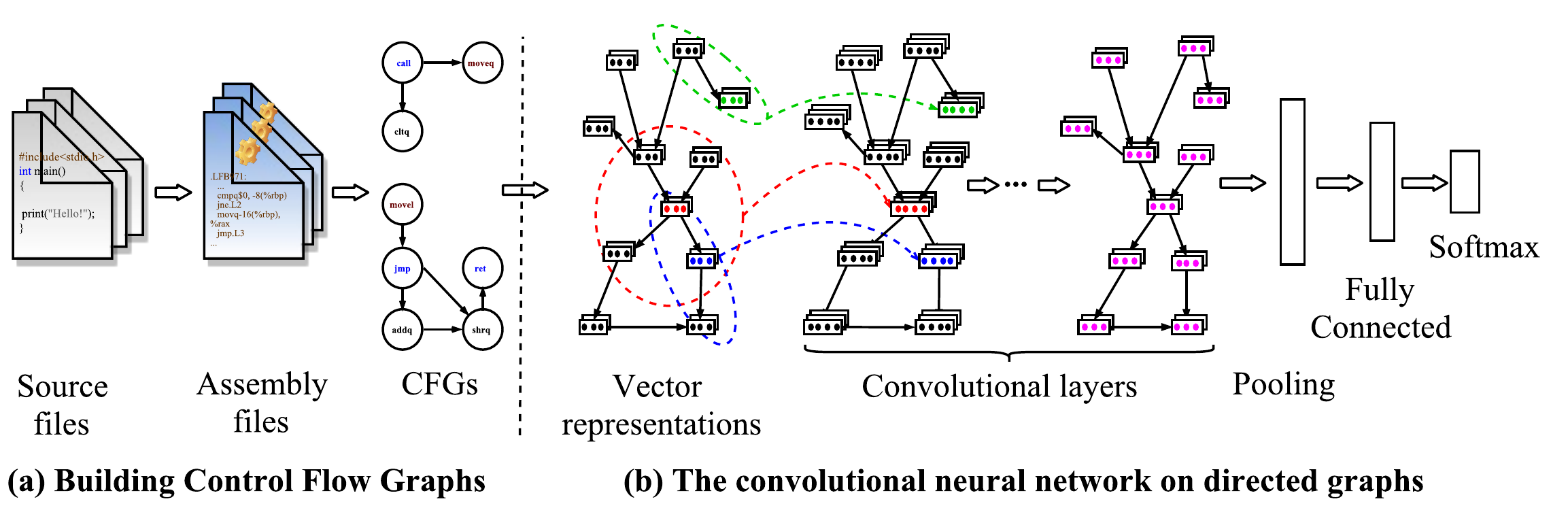}
\caption{The overview of our approaches for software defect prediction using convolutional neural networks on Control Flow Graphs of assembly code.}
\label{fig_gcnn_architecture}
\end{figure*}
The directed graph-based convolutional neural network (DGCNN) is a dynamic graphical model that is designed to treat large-scale graphs with complex information of vertex labels. For instance, a CFG vertex is not simply a token but represents an instruction, which may contain many components including the instruction name and several operands. In addition, each instruction can be viewed in other perspectives (multiple views), e.g. instruction types or functions.

Fig.~\ref{fig_gcnn_architecture} depicts the overview architecture of DGCNN. In the DGCNN model, the first layer is called vector representations or the embedding layer, whereby a vertex is represented as a set of real-valued vectors corresponding with the number of views. Next, we design a set of fixed-size circular windows sliding over the entire graphs to extract local features of substructures. The input graphs are explored at different levels by stacking several convolutional layers on the embedding layer. In our work, we apply DGCNN with two layers in the convolution stage. After convolution, a dynamic pooling layer is applied to gather extracted features from all the parts of the graphs into a vector. Finally, the feature vector is fed into a fully-connected layer and an output layer to compute the categorical distributions for possible outcomes.

Specifically, during the forward pass of convolutional layers, each filter slides through all vertices of the graph and computes dot products between entries of the filter and the input. Suppose that the subgraph in the sliding window includes $d + 1$ vertices (the current vertex and its neighbors) with vector representations of $x_0, x_1, ..., x_d \in \mathbb{R}^{
v_f \times n_f}$, then the output of the filters is computed as follows: 
\begin{equation}
\label{eq_conv_out}
y = tanh(\sum_{i=0}^{d}\sum_{j=1}^{v_f}W_{conv,i,j}\cdot x_{i,j} + b_{conv})
\end{equation}
where $y, b_{conv} \in \mathbb{R}^{v_c \times n_c}, W_{conv,i} \in \mathbb{R}^{v_c \times n_c \times v_f\times n_f}$. $tanh$ is the activation function. $n_f$ and $v_f$ are the vector size and the number of views of the input layer. $n_c$ and $v_c$ are the numbers of filters and views of the convolutional layer.

Due to arbitrary structures of graphs, the numbers of vertices in subgraphs are different. As can be seen in Fig.\ref{fig_gcnn_architecture}, the current receptive field at the red node includes 5 vertices while only 3 vertices are considered if the window moves right down. Consequently, determining the number of weight matrices for filters is unfeasible. To deal with this obstacle, we divide vertices into groups and treat items in each group in a similar way. Regarding the way, the parameters for convolution have only three weight matrices including $W^{cur}, W^{in},$ and $W^{out}$ for current, outgoing, and incoming nodes, respectively.

In the pooling layer, we face the similar problem of dynamic graphs. The numbers of nodes are varying between programs' CFGs. Meanwhile, convolutions preserve the structures of input graphs during feature extraction. This means that it is impossible to determine the number of items which will be pooled into the final feature vector. An efficient solution to this problem is applying dynamic pooling~\cite{socher2011dynamic} to normalize the features such that they have the same dimension. Regarding this, one-way max pooling is adopted to gather the information from all parts of the graph to one fixed size vector regardless of graph shape and size.
\section{Experiments}
\label{sec:experiments}
\subsection {Datasets}
The datasets for conducting experiments are obtained from a popular programming contest site CodeChef \footnote{\url{https://www.codechef.com/problems/<problem-name>}}. We created four benchmark datasets which each one involves source code submissions (written in C, C++, Python, etc.) for solving one of the problems as follows:
\begin{itemize}
\item SUMTRIAN (Sums in a Triangle): Given a lower triangular matrix of $n$ rows, find the longest path among all paths starting from the top towards the base, in which each movement on a part is either directly below or diagonally below to the right. The length of a path is the sums of numbers that appear on that path.
\item FLOW016 (GCD and LCM): Find the greatest common divisor (GCD) and the least common multiple (LCM) of each pair of input integers A and B.
\item MNMX (Minimum Maximum): Given an array $A$ consisting of $N$ distinct integers, find the minimum sum of cost to convert the array into a single element by following operations: select a pair of adjacent integers and remove the larger one of these two. For each operation, the size of the array is decreased by 1. The cost of this operation will be equal to their smaller. 
\item SUBINC (Count Subarrays): Given an array $A$ of $N$ elements, count the number of non-decreasing subarrays of array $A$.
\end{itemize}

The target label of an instance is one of the possibilities of source code assessment. Regarding this, a program can be assigned to one of the groups as follows: 0) accepted - the program ran successfully and gave a correct answer; 1) time limit exceeded - the program was compiled successfully, but it did not stop before the time limit; 2) wrong answer: the program compiled and ran successfully but the output did not match the expected output; 3) runtime error: the code compiled and ran but encountered an error due to reasons such as using too much memory or dividing by zero; 4) syntax error - the code was unable to compile.

We collected all submissions written in C or C++ until March 14th, 2017 of four problems. The data are preprocessed by removing source files which are empty code, and unable to compile. Table~\ref{table:datastatistics} presents statistical figures of instances in each class of the datasets. All of the datasets are imbalanced. Taking MNMX dataset as an example, the ratios of classes 2, 3, 4 to class 0 are 1 to 27, 46, and 24. To conduct experiments, each dataset is randomly split into three folds for training, validation, and testing by ratio 3:1:1.
\begin{table}[]
\centering
\setlength{\tabcolsep}{4pt}
\caption{Statistics on the datasets}
\label{table:datastatistics}
\begin{tabular}{lrrrrrr}
\hline
Dataset & Total & Class 0 & Class 1 & Class 2 & Class 3 & Class 4 \\ \hline
FLOW016 &10648 &3472 &4165 &231 &2368 &412 \\
MNMX    &8745 &5157 &3073 &189 &113 &213 \\
SUBINC  &6484&3263&2685&206&98&232\\
SUMTRIAN&21187&9132&6948&419&2701&1987\\ \hline
\end{tabular}
\end{table}
\subsection {Experimental Setup}
We compare our model with tree-based approaches which have been successfully applied to programming language processing tasks. For the tree-based methods, a source code is represented as an abstract syntax trees (AST) using a parser, and then different machine learning techniques are employed to build predictive models. The settings for baselines are presented as follows:

\begin{table}[]
\centering
\setlength{\tabcolsep}{4pt}
\selectfont
\caption{Structures and numbers of hyperparameters of the neural networks. Each layer is presented in form of the name followed by the number of neurons. Emb is a embedding layer. Rv, TC, GC, and FC stand for recursive, tree-based convolutional, graph-based convolutional, and fully-connected, respectively.}
\label{table_netstructures}
\newcommand{\specialcell}[2][c]{%
  \begin{tabular}[#1]{@{}c@{}}#2\end{tabular}}
\begin{tabular}{llcc}
\hline
Network & \multicolumn{1}{c}{Architecture}& weights& biases\\
\hline
RvNN & Coding30-Emb30-Rv600-FC600-Soft5&1104600&1235\\
TBCNN & Coding30-Emb30-TC600-FC600-Soft5&1140600&1235\\
SibStCNN & Coding30-Emb30-TC600-FC600-Soft5&1140600&1235\\
DGCNN-1V & GC100-GC600-FC600-Soft5&552000&1305\\
DGCNN-2V & GC100-GC600-FC600-Soft5&561000&1305\\
\hline
\end{tabular}
\end{table}

\textbf{The neural networks} For neural networks including DGCNN, tree based convolutional neural networks (TBCNN)~\cite{mou2016convolutional}, Sibling-subtree convolutional neural networks (SibStCNN - an extension of TBCNN in which feature detectors are redesigned to cover subtrees including a node, its descendants and siblings), and recursive neural networks (RvNN), we use some common hyper-parameters: initial learning rate is 0.1, vector size of tokens is 30. The structures of the networks are shown in Table~\ref{table_netstructures}. 

When adapting DGCNN for CFGs, we use two views of CFG nodes including instructions and their type. For instance, \texttt{jne}, \texttt{jle}, and \texttt{jge} are assigned into the same group because they are conditional jump instructions. For instructions, it should be noted that they may have many operands. In this case, we replace all operands such as block names, processor register names, and literal values with symbols ``name", ``reg'', and ``val'', respectively. Following this replacement, the instruction \texttt{addq \$32, \%rsp} is converted into \texttt{addq value, reg}. 

To generate inputs for DGCNN, firstly the symbol vectors are randomly initialized and then the corresponding vector of each view is computed based on the component vectors as follows:
\begin{equation}
\label{eq:vec_of_view}
x_{v_i} = \frac{1}{C}\sum_{j=1}^{C}{x_j}
\end{equation}
where $C$ is the number of symbols in the view, and $x_j$ is the vector representation of the $j^{th}$ symbol. Taking instruction \texttt{addq \$32, \%rsp} as an example, its vector is the linear combination of vectors of symbols \texttt{addq}, \texttt{value}, and \texttt{reg} 

\textbf{k-nearest neighbors (kNN)} We apply kNN algorithm with tree edit distance (TED) and Levenshtein distance (LD)~\cite{phan2016exploiting}. The number of neighbors $k$ is set to 3. 

\textbf{Support Vector Machines (SVMs)} The SVM classifiers are built based on hand-crafted features, namely bag-of-words (BoW). By this way, the feature vector of each program is determined by counting the numbers of times symbols appear in the AST. The SVM with RBF kernel has two parameters $C$ and $\gamma$; their values are 1 and 0, respectively. 
\subsection {Evaluation Measures}
The approaches are evaluated based on two widely used measures including accuracy and the area under the receiver operating characteristic (ROC) curve, known as the AUC. 

Predictive accuracy has been considered the most important criterion to evaluate the effectiveness of classification algorithms. For multi-class classification, the accuracy is estimated by the average hit rate.

In machine learning, the AUC that estimates the discrimination ability between classes is an important measure to judge the effectiveness of algorithms. It is equivalent to the non-parametric Wilcoxon test in ranking classifiers~\cite{fawcett2006introduction}. According to previous research, AUC has been proved as a better and more statistically consistent criterion than the accuracy~\cite{ling2003auc}, especially for imbalanced data. In the cases of imbalanced datasets that some classes have much more samples than others, most of the standard algorithms are biased towards the major classes and ignore the minor classes. Consequently, the hit rates on minor classes are very low, although the overall accuracy may be high. Meanwhile, in practical applications, accurately predicting minority samples may be more important. Taking account of software defect prediction, the essential task is detecting faulty modules. However, many software defect datasets are highly imbalanced and the faulty instances belong to minority classes~\cite{rodriguez2014preliminary}. Because all experimental datasets are imbalanced, we adopt both measures to evaluate the classifiers.

ROC curves which depict the tradeoffs between hit rates and false alarm rates are commonly used for analyzing binary classifiers. To extend the use of ROC curves to multi-class problems, the average results are computed based on two ways: 1) macro-averaging gives equal weight to each class, and 2) micro-averaging gives equal weight to the decision of each sample~\cite{sokolova2009systematic}. The AUC measure for ranking classifiers is estimated by the area under the macro-averaged ROC curves.

\section{Results and Discussion}
\label{sec:results}
Table~\ref{table:result_acc} shows the accuracies of classifiers on the four datasets. As can be seen, CFG-based approaches significantly outperform others. Specifically, in comparison with the second best, they improve the accuracies by 12.39\% on FLOW016, 1.2\% on MNMX, 7.71\% on SUBINC, and 1.98\% on SUMTRIAN. As mentioned before (Section~\ref{sec:introduction}), software defect prediction is a complicated task because semantic errors are hidden deeply in source code. Even if a defect exists in a program, it is only revealed during running the application under specific conditions. Therefore, it is impractical to manually design a set of good features which are able to distinguish faulty and non-faulty samples. Similarly, ASTs just represent the structures of source code. Although tree-based approaches (SibStCNN, TBCNN, and RvNN) are successfully applied to other software engineering tasks like classifying programs by functionalities, they have not shown good performance on the software defect prediction. In contrast, CFGs of assembly code is precise graphical structures which show behaviors of programs. As a result, applying DGCNN on CFGs achieves the highest accuracies on the experimental datasets about software defects.
\begin{table}[]
\centering
\setlength{\tabcolsep}{5pt}
\caption{Comparison of classifiers according to accuracy. 1V and 2V following DGCNN means that CFG nodes are viewed by one and two perspectives. Op and NoOp are using instruction with or without operands.}
\label{table:result_acc}
\begin{tabular}{lcccc}
\hline
Approach   & FLOW016 & MNMX   & SUBINC & SUMTRIAN \\ \hline
SVM-BoW     &  60.00 &  77.53 &  67.23 &  64.87   \\
LD          &  60.75 &  79.13 &  66.62 &  65.81   \\
TED         &  61.69 &  80.73 &  68.31$^*$ &  66.97$^*$ \\
RvNN        &  61.03 &  82.56 &  64.53 &  58.82   \\
TBCNN       &  63.10$^*$ &  82.45 &  63.99 &  65.05   \\
SibStCNN    &  62.25 &  82.85$^*$ &  67.69 &  65.10   \\ \hline
DGCNN\_1V\_NoOp    &  73.80 &  83.19 &  70.93 &  68.83   \\
DGCNN\_2V\_NoOp    &  74.32 &  83.82 &  74.02 &  68.12    \\
DGCNN\_1V\_Op &  \textbf{75.49} &  \textbf{84.05} &  72.40 &  68.19   \\
DGCNN\_2V\_Op &  75.12 &  83.71 &  \textbf{76.02} &  \textbf{68.95}   \\\hline      
\end{tabular}
\end{table}

From the last four rows of Table~\ref{table:result_acc}, the more the information is provided, the more efficient the learner is. In general, viewing graph nodes by two perspectives including instructions and instruction groups helps boost DGCNN classifiers in both cases: with and without the use of operands. Similarly, taking into account of all components in instructions (Eq.~\ref{eq:vec_of_view}) is beneficial. In this case, the DGCNN models achieve highest accuracies on the experimental datasets. Specifically, DGCNN with one view reaches the accuracies of 75.49\% on FLOW016, and 84.05\% on MNMX; DGCNN with two views obtains the accuracies of 76.02\% on SUBINC, and 68.95\% on SUMTRIAN.  

\begin{figure}[]
  \centering
   \vfill \vspace{-0.4cm}
  \subfloat[]{\includegraphics[width=8.5cm]{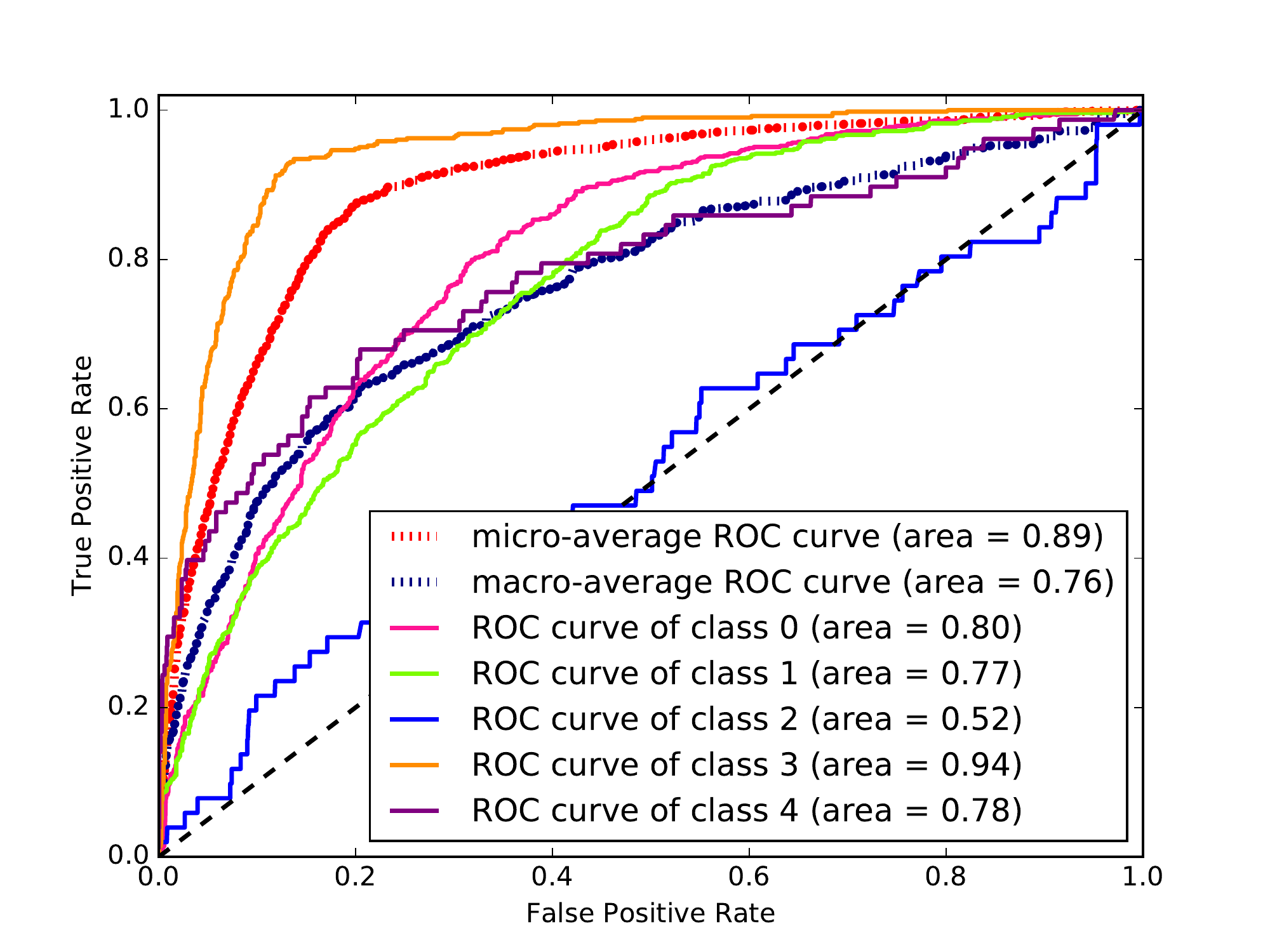}\label{fig:FLOW016_TBCNN_roc}}
  \vfill \vspace{-0.4cm}
  \subfloat[]{\includegraphics[width=8.5cm]{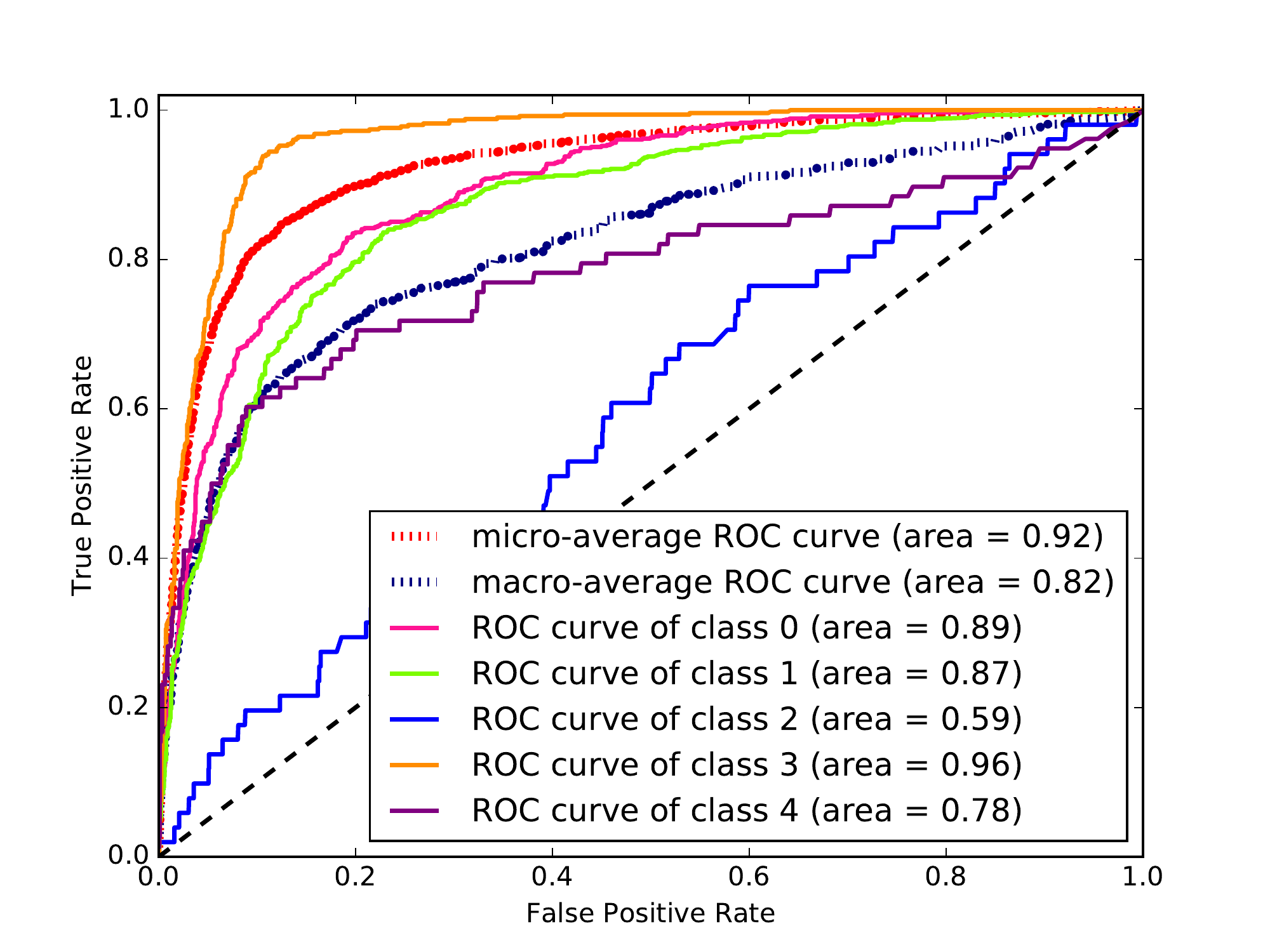}\label{fig:FLOW016_DGCNN_1V_NoOp_roc}}
  \caption{The illustration of the discrimination ability between classes of classifiers on imbalanced datasets. Fig.~\ref{fig:FLOW016_TBCNN_roc} and Fig.~\ref{fig:FLOW016_DGCNN_1V_NoOp_roc} are the ROC curves of TBCNN and DGCNN\_1V\_NoOp on MNMX dataset, respectively.}
 \label{fig:classifier_rocs_onMNMX}
\end{figure}

We also assess the effectiveness of the models in terms of the discrimination measure (AUC) which is equivalent to Wilcoxon  test in ranking classifiers. For imbalanced datasets, many learning algorithms have a trend to bias the majority class due to the objective of error minimization. As a result, the models mostly predict an unseen sample as an instance of the majority classes, and ignore the minority classes. Fig.~\ref{fig:classifier_rocs_onMNMX} plots the ROC curves of TBCNN and DGCNN\_1V\_NoOp classifiers on MNMX dataset, an imbalanced data with the minority classes of 2 and 4. Both two classifiers have a notable lower ability in detecting minority instances from the others. For predicting class 4, the TBCNN is even equivalent to a random classifier. After observing the other ROC curves we found the similar problem for all of the approaches on the experimental datasets. Thus, AUC is an essential measure for evaluating classification algorithms, especially in the case of imbalanced data.  
\begin{table}[]
\centering
\setlength{\tabcolsep}{5pt}
\caption{The performance of classifiers in terms of AUC.}
\label{table:result_AUC}
\begin{tabular}{lcccc}
\hline
Approach    & FLOW016 & MNMX   & SUBINC & SUMTRIAN \\ \hline
SVM-BoW     & 0.74  & 0.76  & 0.73  &  0.79\\
RvNN        & 0.75  & 0.79  & 0.69  & 0.73\\
TBCNN       & 0.76  & 0.77  & 0.72 &  0.78 \\
SibStCNN    & 0.76  & 0.79  & 0.71 &  0.80 \\ \hline
DGCNN\_1V\_NoOp    & \textbf{0.82}  & \textbf{0.82}  & 0.74  & \textbf{0.82}  \\
DGCNN\_2V\_NoOp    & 0.80  & 0.81 & 0.72 &  0.81\\
DGCNN\_1V\_Op & 0.81 & 0.80 & \textbf{0.75}&  0.81 \\
DGCNN\_2V\_Op & \textbf{0.82} & 0.79  & 0.74 &  0.81 \\ \hline      
\end{tabular}
\end{table}

Table~\ref{table:result_AUC} presents the AUCs of probabilistic classifiers, which produce the probabilities or the scores to indicate the belonging degrees of an instance to classes. There are two groups including graph-based and tree-based approaches, in which the approaches in each group has the similar AUC scores; and graph-based approaches show better performance than those of tree-based. It is worth noticing that, along with the efforts of accuracy maximization, the approach based on DGCNN and CFGs also enhance the distinguishing ability between categories even on imbalanced data. The DGCNN classifier improves the second best an average of 0.03 on AUC scores. From above analysis, we can conclude that leveraging precise control flow graphs of binary codes is suitable for software defect prediction, one of the most difficult tasks in the field of software engineering. 
\section{Error Analysis}
\label{sec:error_analysis}
\begin{figure*}
  \centering
  \subfloat[File 3.c (a training sample)]{\includegraphics[width=6cm,valign=t]{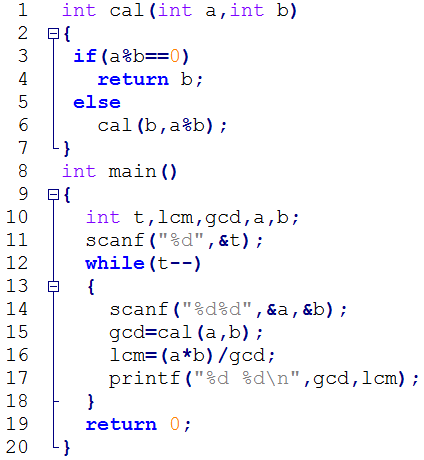}\label{fig:code_training}}
  \hfill
  \subfloat[ File 4.c (\textbf{G+, T-})]{\includegraphics[width=7.9cm,valign=t]{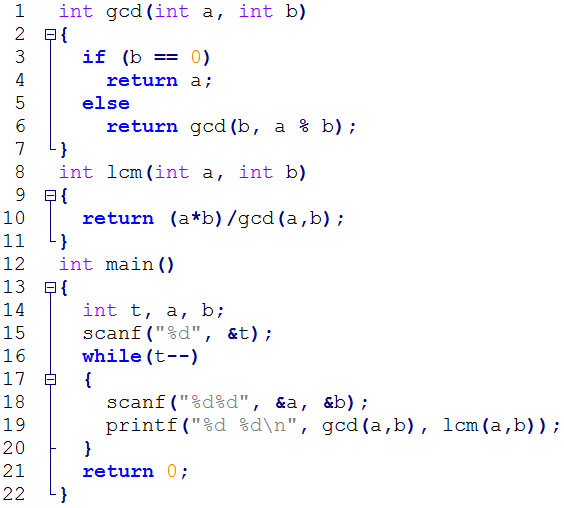}\label{fig:code_gcnntrue_sibfalse}}
  \vfill
  \subfloat[File 5.c (\textbf{G-, T+})]{\includegraphics[width=6.74cm,valign=t]{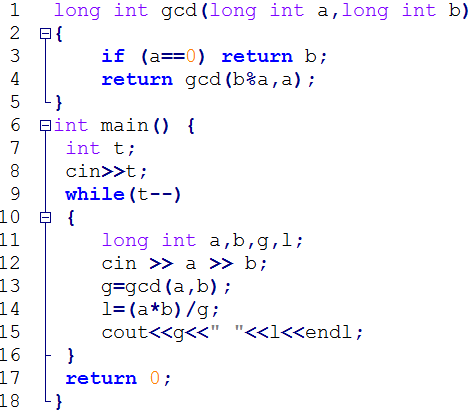}\label{fig:code_gcnnfalse_sibtrue}}
   \hfill
  \subfloat[File 6.c (\textbf{G-, T-})]{\includegraphics[width=7.38cm,valign=t]{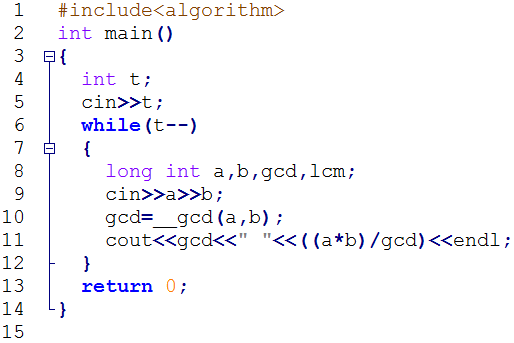}\label{fig:code_gcnnfalse_sibfalse}}
  \caption{Some source code examples in FLOW016 dataset which may cause mistakes of tree-based (T) and CFG-based (G) approaches. Fig.~\ref{fig:code_training} is a sample in the training set. Figs.~\ref{fig:code_gcnntrue_sibfalse},~\ref{fig:code_gcnnfalse_sibtrue}, and~\ref{fig:code_gcnnfalse_sibfalse} are samples in the test set. Symbols ``+'' and ``-'' denote the sample is correctly and incorrectly classified by the approaches.}
  \label{fig:code_examplesofmistakes}
\end{figure*}
We analyze cases of source code variations which methods are able to handle or not based on observations on classifiers' outputs, training and test data. We found that RvNN's performance is degraded when tree sizes increase. This problem is also pointed out from other research on tasks of natural language processing~\cite{socher2011semi,socher2013recursive} and programming language processing~\cite{mou2016convolutional}. From Tables~\ref{table:datastatistics}, ~\ref{table:result_acc}, and ~\ref{table:result_AUC} the larger trees, the lower accuracies and AUCs, RvNN obtains in comparison with other approaches, especially on SUMTRIAN dataset. SibStCNN and TBCNN obtain higher performance than other baselines due to learning features from subtrees. For analyzing tree-based methods in this section, we only take into account SibStCNN and TBCNN. 

\textbf{Effect of code structures}: the tree-based approaches suffer from varying structures of ASTs. For example, given a program, we have many ways to reorganize the source code such as changing positions of some statements, constructing procedures and replacing statements by equivalent ones. These modifications lead to reordering the branches and producing new branches of ASTs (\texttt{File 3.c} and \texttt{File 4.c}). Because of the weight matrices for each node being determined based on the position, SibStCNN and TBCNN are easily affected by changes regarding tree shape and size. 

Meanwhile, graph-based approaches are able to handle these changes. We observed that although loop statements like \texttt{For}, \texttt{While}, and \texttt{DoWhile} have different tree representations, their assembly instructions are similar by using a jump instruction to control the loop. Similarly, moving a statement to possible positions may not result in notable changes in assembly code. Moreover, grouping a set of statements to form a procedure is also captured in CFGs by using edges to simulate the procedure invocation (Section~\ref{subsec:CFG}).         

\textbf{Effect of changing statements}: CFG-based approaches may be affected by replacements of statements. Considering source code in \texttt{File 3.c}, \texttt{File 5.c}, they have similar ASTs, but the assembly codes are different. In \texttt{C} language, statements are translated into different sets of assembly instructions. For example, with the same operator, the sets of instructions for manipulating data types of \texttt{int} and \texttt{long int} are dissimilar. Moreover, statements are possible replaced by others without any changes of program outcomes. Indeed, to show values, we can select either \texttt{printf} or \texttt{cout}. Since contents of CFG nodes are changed significantly, DGCNN may fail in predicting these types of variations.  

\textbf{Effect of using library procedures}: when writing a source code, the programmer can use procedures from other libraries. In Fig.~\ref{fig:code_examplesofmistakes}, \texttt{File 6.c} applies the procedure \texttt{\_\_gcd} in the library \texttt{algorithm}, while the others use ordinary C statements for computing the greatest common divisor of each integer pair. Both ASTs and CFGs do not contain the contents of external procedures because they are not embedded to generate assembly code from source code. As a result, tree-based and graph-based approaches are not successful in capturing program semantics in these cases.        
\section{Conclusion}
\label{sec:conclusions}
This paper presents an end-to-end model for solving software defect prediction, one of the most difficult tasks in the field of software engineering. By applying precise representations (CFGs) and a graphical deep neural network, the model explores deeply the behavior of programs to detect faulty source code from others. Specifically, the CFG of a program is constructed from the assembly code after compiling its source code. Then DGCNN is leveraged to learn from various information of CFGs data to build predictive models.

Our evaluation of four real-world datasets indicates learning on graphs could significantly improve the performance of feature-based and tree-based approaches according to both accuracy and discrimination measures. Our method improves the accuracies from 4.08\% to 15.49\% in comparison with the feature-based approach, and from  1.2\% to 12.39\% in comparison with the tree-based approaches.
\section*{Acknowledgements}
This work was supported partly by JSPS KAKENHI Grant number 15K16048 and the first author would like to thank the scholarship from Ministry of Training and Education (MOET), Vietnam under the project 911.
\bibliographystyle{plain}
\bibliography{ref}
\end{document}